# A new time quantifiable Monte Carlo method in simulating magnetization reversal process

X. Z. Cheng and M. B. A. Jalil

*Department of Electrical and Computer Engineering, National University of Singapore, 4 Engineering Drive 3,*

*Singapore 117576*

H. K. Lee and Y. Okabe

*Department of Physics, Tokyo Metropolitan University, 1-1 Minami-Osawa, Hachioji-shi, Tokyo, Japan 192-0397*

We propose a new time quantifiable Monte Carlo (MC) method to simulate the thermally induced magnetization reversal for an isolated single domain particle system. The MC method involves the determination of density of states, and the use of Master equation for time evolution. We derive an analytical factor to convert MC steps into real time intervals. Unlike a previous time quantified MC method, our method is readily scalable to arbitrarily long time scales, and can be repeated for different temperatures with minimal computational effort. Based on the conversion factor, we are able to make a direct comparison between the results obtained from MC and Langevin dynamics methods, and find excellent agreement between them. An analytical formula for the magnetization reversal time is also derived, which agrees very well with both numerical Langevin and time-quantified MC results, over a large temperature range and for parallel and oblique easy axis orientations.

75.40.Gb, 75.40.Mg, 75.50.Tt

## I. INTRODUCTION

Several simulation methods based on the Monte Carlo (MC) approach have been used to study the thermally-induced magnetization reversal of magnetic particles. These include the kinetic MC method,[1] which assumes that the system resides only in the energy minima states, and that the transition rate over the energy barrier $\Delta E$ separating the two minima obeys the Arrhenius-Neel's law. The characteristic time constant in this method, however is available for a few simple cases only.[2] Other MC techniques such as the standard Metropolis algorithm[3], absorbing Markov Chains algorithm[4] and Projection MC method[5] are also helpful in describing equilibrium properties and relaxation processes over long time scales for complex systems. However, a major disadvantage of these MC simulation techniques is that time is calibrated in terms of Monte Carlo steps (MCS). Unfortunately, the conversion of MCS into real (physical) time units is not a trivial problem.

Nowak *et al.*[6] first proposed an analytical time quantification of the Metropolis Monte Carlo method applied to isolated single domain magnetic particles. The accuracy of the time quantification is confirmed by a comparison with numerical Langevin dynamics (LD) results. The LD approach is an alternative stochastic approach for modeling thermally induced magnetization reversal. This method involves the numerical integration of the stochastic Landau-Lifshitz-Gilbert (LLG) dynamical equation of motion. The LLG equation governs the time evolution of the particle magnetization **M** and incorporates the precession, damping and thermal fluctuations of **M**. As proposed by Brown[7], the effect of thermal fluctuations is incorporated as a randomly oriented white noise field contribution to the total effective field. Unlike the MC method, the LD method is calibrated in real physical time. However, it is usually suitable for modeling short time-scale dynamics because the maximum time step size $\Delta t$ is only of the order of several ps. The upper limit of $\Delta t$ is constrained by the reciprocal of the gyromagnetic constant $g_0$, which typically of the order $10^7$ Hz/Oe in common magnetic materials [e.g. Ref. (8)]. It is thus technically infeasible to perform a LD integration much beyond a time scale of a few ns.

In Ref. (6), Nowak *et al.* achieved the time quantification relationship of the metropolis Monte Carlo method by deriving the analytical relation between the MC step size and the mean squared deviation of the magnetic moment orientation. Chubykalo *et al.*[9] investigated the constraints on the validity of Nowak's conversion scheme, especially with regards to athermal (energy conserving) precessional motion. Further research works have also yielded proof of the validity of Nowak's time quantification relationship in a coupled nanomagnetic particle array system.[10]

In this article, we present another time quantifiable Monte Carlo method in simulating the magnetization reversal process. Our model applies the Wang-Landau random walk Monte-Carlo (RWMC) algorithm[11] to determine the density of states $g(E, M)$ as a function of energy and magnetization. The Wang-Landau algorithm was chosen because of its greater efficiency compared to other numerical methods of calculating the density of states, e.g. multicanonical methods[12,13], flat histogram method[14,15] and broad histogram method[16]. From $g(E,M)$, a tridiagonal transition matrix is obtained by applying Glauber transition rate, and the resulting Master equation solved explicitly. This method was first applied by Lee *et al.* for an array of Ising spins[17]. In this article, we extend it to the Heisenberg model (3D continuous spin orientation) for a single isolated particle. The main result of our work is the time quantification of this MC method, which is achieved by approximating the discrete Master equation into the corresponding (continuous) Fokker Planck (FP) equation in the limit of large bin number. This FP equation forms a critical bridge to the Langevin dynamics method, for which the FP diffusion term (related to thermal fluctuations) is well-known. Comparing the diffusive FP terms for both MC and LD methods, we obtain an *analytical* conversion factor for MCS into real time steps. The conversion factor is validated by performing numerical MC and LD simulations. We achieve very good convergence between the LD and time-quantified MC data over a wide range of temperatures. As an independent check, the converged LD and time-quantified MC results for the parallel easy-axis case, also show good agreement with Brown's asymptotic prediction[18] at low temperature.

Compared to the time-quantified Metropolis Monte Carlo method of Nowak *et al*, our time quantified MC method has two main advantages. First, the density of states $g(E,M)$ is independent of temperature $T$, which means that for a given system, its equilibrium state $P_{eq}$ can be analytically derived for any $T$ once $g(E,M)$ is known. Second, the state of the system can be obtained at any time $t$ once $P_{eq}$ and the eigenfunctions of the rate matrix are known. Hence, the relaxation process can be modeled for arbitrarily long time duration and at any arbitrary $T$, without any increase in computational effort. By contrast, in the Metropolis scheme, the computational time increases linearly with $t$, and the time-consuming stochastic MC modeling has to be repeated to model the switching behavior at different $T$.

Our final result concerns the derivation of the switching time $t$, expressed as an function of $g(E,M)$. This derivation is performed within the MC framework, and coupled with the time quantification, it allows us to obtain an *analytical* estimate of the reversal time in real physical time units without the need for any numerical simulation, once $g(E,M)$ is obtained from the RWMC algorithm. This result is applicable to any easy-axis orientation and at any arbitrary temperature. Unlike simpler approximations based on the second-largest eigenvalue $l_1$ of the rate matrix, our refined expression takes into accounts the contribution of all eigenfunctions, and show a much closer agreement to the numerical simulation results.

## II. MODEL AND METHODS

The system under consideration is an isolated Brownian single domain particle. The free energy of the particle in the Heisenberg model consists of anisotropy and Zeeman energy energy, i.e.

$$E_{tot} = -K_u V(\mathbf{S} \cdot \mathbf{k}_n)^2 - m_s \mathbf{B} \cdot \mathbf{S} \quad (1)$$

where $K_u$ is the anisotropy constant, $\mathbf{S} = \mathbf{M}/M_s$ is the normalized magnetization and $\mathbf{k_n}$ the unit vector along the easy axis direction. In Eq. (1), the z axis is chosen to correspond to the external field's direction.

**Langevin dynamics**

The Langevin dynamics of the magnetization $\mathbf{S}$ is described in the form of a reduced LLG equation:

$$\frac{d\mathbf{S}}{dt} = -\mathbf{S} \times (\mathbf{h}_{eff} + \mathbf{h}(t)) - a \cdot \mathbf{S} \times (\mathbf{S} \times (\mathbf{h}_{eff} + \mathbf{h}(t)))$$
(2)

The normalized dimensionless variables are defined as $\mathbf{h}_{eff} = (\mathbf{H}_{eff}/H_k)$ and $t = g_0 H_k t/(1+a^2)$ where $H_k = (2K_u/m_0 M_s)$. The effective field is obtained from Eq. (1) i.e. $\mathbf{H}_{eff} = -(\partial E_{tot}/\partial \mathbf{M}) = -M_s^{-1}(\partial E_{tot}/\partial \mathbf{S})$. In the above, $g_0$ represents the gyromagnetic ratio and $a$ the damping constant. $\mathbf{h}(t)$ is the additional field acting on $\mathbf{S}$ due to thermal effects, and is represented by a white noise term with the following statistical properties:[7]

$$\langle \mathbf{h}(t) \rangle = 0,$$

$$\langle h_i(0) \cdot h_j(t) \rangle = \frac{\boldsymbol{a} \cdot k_B T}{(1+\boldsymbol{a}^2) K_u V} \boldsymbol{d}_{ij} \boldsymbol{d}(t), \quad (3)$$

where $i, j$ denote Cartesian components $x, y, z$. In this work, the numerical integration of the LLG equation [Eq. (2)] is done via the Heun scheme, using the Stratonovich interpretation, with the reduced time interval set at $\Delta t = 0.01$, which is sufficiently small to ensure stability.

**Monte-Carlo method**

Monte Carlo methods have been used to study the dynamics of magnetic reversal of a system with meta-stable states [19]. In general, we can write down the corresponding microscopic master equation

$$\frac{dP(M,t)}{dt} = \sum_{M'} w(M|M') P(M',t) - w(M'|M) P(M,t) \quad (4)$$

where $M$ and $M'$ are magnetization states and $w(M|M')$ is the transition rate from state $M'$ to state $M$.

Due to the physical grounds that magnetization transitions are continuous in the limit of small time step, a reasonable approximation is to restrict the transitions to occur between adjacent states only. Thus, Eq. (4) serves as a description of a diffusion process. Lee *et al.* first developed this Master equation method combined with random walk MC algorithm to solve magnetization reversal process of interacting Ising spin arrays and discussed circumstances for its validity[17]. Here, the method is extended to model the Heisenberg system by binning the magnetization orientation into a finite number $N$ of discrete values. With the restriction of transitions between adjacent states only, Eq. (4) can be written in a matrix form as

$$\frac{d\vec{P}(t)}{dt} = \mathbf{A} \cdot \vec{P}(t) \quad (5)$$

where the transition matrix

$$\mathbf{A}_{n \times n} = \begin{pmatrix} -w_1 & u_1 & & & \\ w_1 & -w_2-u_1 & u_2 & & \\ & w_2 & \cdots & & \\ & & & \cdots & u_{n-1} \\ & & & w_{n-1} & -u_{n-1} \end{pmatrix} \quad (6)$$

In the above matrix, $w_i = w(M_{i+1}|M_i)$ and $u_i = w(M_i|M_{i+1})$. Various simulation methods to approximate $w(M|M')$ have been discussed previously, including mean field dynamics[20] and Transition Matrix Monte Carlo[21] methods. In our simulation, we assume the Glauber transition rate i.e.

$$w(M|M') = 1/(1 + P_{eq}(M')/P_{eq}(M)) \quad (7)$$

Where $P_{eq}(M)$ is the equilibrium (stationary) probability distribution function of state $M$. $P_{eq}(M)$ is estimated based on the density of states $g(E,M)$ obtained using the Wang-Landau random walk algorithm[11,17].

$$Z = \sum_{E,M} g(E,M) \exp(-\boldsymbol{b}E)$$
$$P_{eq}(M) = g(E,M) \cdot \exp(-\boldsymbol{b}E)/Z \quad (8)$$

Note that for the case of an isolated single domain particle system with easy axis parallel to the applied field, the (analytical) density of states function is known i.e. $g(E,m) = k$ (normalized constant). However, we have used the Wang-Landau random walk algorithm together with the Master equation solution as a complete MC method, which is applicable to the more general case of oblique easy axis directions, and scalable to Heisenberg spin array systems. The explicit solution of Eq (5) can be expressed as an eigenvalue expansion[17,22].

$$\vec{P}(t) = \vec{P}_{eq} + \sum_{i=1}^{N-1} \boldsymbol{a}_i \vec{v}_i \exp(\boldsymbol{l}_i t) \quad (9)$$

where $\vec{P}_{eq}$ is the equilibrium probability distribution function in vector form, and $\boldsymbol{l}_i$ and $\vec{v}_i$ are non-zero eigenvalues and corresponding eigenvectors of the transition matrix $\mathbf{A}$. The factors $\boldsymbol{a}_i$ are determined based on the initial conditions. Thus, once $\boldsymbol{l}_i$ and $\vec{v}_i$ are known, Eq. (9) will yield the magnetization probability distribution at any arbitrary time $t$.

The solution of the full set of $\boldsymbol{l}_i$ and $\vec{v}_i$ can only be obtained numerically. However, one can derive some analytical results based on the property of the transition matrix $\mathbf{A}$ as expressed in Eq. (6). The eigenvalues of $\mathbf{A}$ are all negative with one exception, which is a zero eigenvalue corresponding to the stationary probability distribution, i.e. $\boldsymbol{l}_0 = 0$ and $\vec{v}_0 = \vec{P}_{eq}$. Due to the exponential time dependence in Eq. (9), it is the second largest eigenvalue $\boldsymbol{l}_1$ (i.e. largest apart from $\boldsymbol{l}_0$) which controls the reversal process. Thus the relaxation time $\boldsymbol{t}_{MC}$ can reasonably be approximated as $\boldsymbol{t}_{MC} = |\boldsymbol{l}_1^{-1}|$.

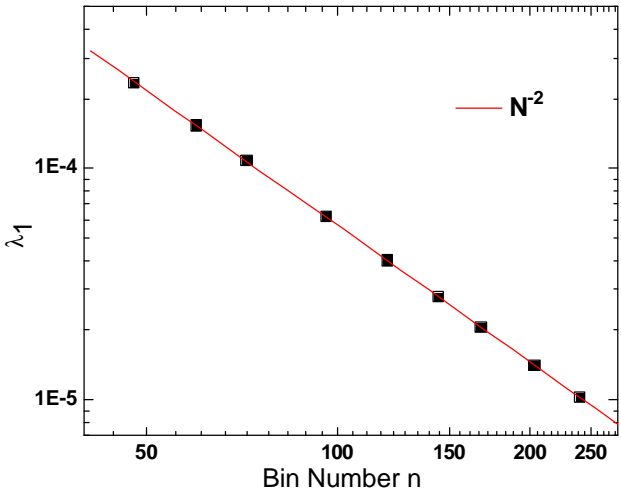

Fig. 1. $|l_1|$ as a function of bin number N. The solid line is a linear fit, which yields a $l_1 \propto N^{-2}$ dependence.

The largest non-zero eigenvalue $|l_1|$ of **A** is determined numerically for different bin number $N$ and plotted in Fig. 1. A very close dependence of $l_1 \propto N^{-2}$ is obtained over the range of bin size considered. This dependence can be reasonable understood by treating the thermally induced magnetization dynamics as a diffusion process where the diffusion length scales as $L \propto t_{dif}^{1/2}$ ($t_{dif}$ being the diffusion time).

The exact solution of $l_1$, however, involves a long analytical expression for the root of an $N^{th}$ order polynomial equation. Furthermore, an approximation based only on $l_1$ will rapidly lose its accuracy at high temperature, i.e. when the contributions of eigenvalues other than $l_1$ become significant. A more useful analytical estimate which we term as the effective eigenvalue $l_{eff}$ can be derived based on a first order approximation (the full derivation is attached in the appendix). This estimate takes into account the contribution of all eigenvalues, and is given by

$$t_{MC} = \left|\frac{1}{l_{eff}}\right| = \sum_{i=1}^{n-1}\left(\left(\sum_{j=1}^{i} p_j\right)\left(\sum_{k=i+1}^{n} p_k\right)\left(\frac{1}{p_i}+\frac{1}{p_{i+1}}\right)\right) \quad (10)$$

where $p_j = P_{eq}(M_j)\big/\sum_i P_{eq}(M_i)$ is the normalized equilibrium probability distribution of state $j$. We will show later that $t_{MC} = l_{eff}^{-1}$ achieves a much closer agreement with the numerical switching time obtained from LD results, compared to the simpler $t_{MC} = |l_1^{-1}|$ estimate, due to the inclusion of the contribution from other eigenvalues, e.g. $l_2, l_3$ etc. The accuracy of $t_{MC} = |l_{eff}^{-1}|$ exceeds that of $|l_1^{-1}|$ especially when the condition $|l_1| \ll |l_j|$ for $j = 2,...,N-1$ is no longer valid.

### III. TIME QUANTIFICATION OF MCS

From a statistical point of view, both LD and MC methods are methods which describe a diffusion process. In the limit of a short time step $\Delta t$, the Langevin dynamics of a Brownian particle can approximately be described by a master equation[23]. During a small time step $\Delta t$ in LD integration, the magnetization $x$ of a macroscopic Brownian particle can only change by a small amount. It is thus reasonable to assume that the magnetization transition rate is appreciable only when the change in magnetization $x = |\Delta x|$ is sufficiently small[17,23]. In this way, the Master equation for the LD method can be obtained which is equivalent to the Master equation (5) of the MC method, in the limit of large bin number in the latter (analogous to the condition for small $x$ in the LD method).

The equivalence of the two methods enables us to obtain the time quantification of the MC method. To determine the time conversion factor from MCS in MC method to real time in LD method, the bridging equation is the Fokker Planck equation (time differential equation describing the probability distribution of a system) which can describe both MC and LD methods, as we shall show below. The general form of Fokker Planck equation is given by

$$\frac{\partial W}{\partial t} = -\frac{\partial}{\partial x}(A \cdot W) + \frac{1}{2}\frac{\partial^2}{\partial x^2}(B \cdot W) \quad (11)$$

where $x \in [-1,1]$ is the normalized magnetization in the field direction (z axis) and $W(q,f,t)$ is the probability distribution function, and is reducible to $W(x,t)$ for the case of uniaxial anisotropy. $A$ and $B$ are the so-called drift and diffusion coefficients respectively, and their values are defined by[23,24]:

$$A = \lim_{\Delta t \to 0} \frac{1}{\Delta t}\langle \Delta x \rangle, \quad B = \lim_{\Delta t \to 0}\frac{1}{\Delta t}\langle \Delta x^2 \rangle. \quad (12)$$

In the Langevin dynamics scheme as discussed by Brown[7], the thermal agitation term in the LLG equation acts as a diffusion term. This term corresponds to coefficient $B$ in the equivalent Fokker Planck equation, and tends to spread out the probability distribution of the spin vector orientation. For an isolated single domain particle undergoing a thermally induced magnetization reversal process, the coefficient $B$ reflects the thermal influence on the magnetization reversal. It can be calculated from the Langevin dynamics scheme in the limit of high damping, and is given by[6,7]

$$B_{LLG} = \frac{2k_B T a g_0}{(1+a^2)m_s} \quad (13)$$

In the MC scheme, when magnetization bin size $x$ is sufficiently small, we can convert the master equation into a continuous differential equation. Considering the $i^{th}$ magnetization state of Eqs. (5) and (6), we thus have

$$\begin{aligned}\frac{dp_i}{dt} &= w_{i-1}p_{i-1} - w_i p_i + u_i p_{i+1} - u_{i-1}p_i \\ &= \left(\exp(-x \cdot \partial/\partial x)-1\right)w_i p_i \\ &\quad + \left(\exp(x \cdot \partial/\partial x)-1\right)u_{i-1}p_i \\ &= \sum_{n=1}^{\infty}(-\partial/\partial x)^n D_i^{(n)} p_i\end{aligned} \quad (14)$$

where the coefficients in the last equality are given by:

$$D_i^{(n)} = (x^n/n!)[w_i + (-1)^n u_{i-1}] \quad (15)$$

Omitting the higher order ($n>2$) expansion in Eq. (14), we can thus rewrite Eq. (5) into the following:

$$\frac{d\bar{p}}{dt} = \frac{d}{dx}\mathbf{A}_{MC}\cdot\bar{p} + \frac{1}{2}\frac{d^2}{dx^2}\mathbf{B}_{MC}\cdot\bar{p} \quad (16)$$

where $\mathbf{A}_{MC}$ and $\mathbf{B}_{MC}$ are diagonal matrices with

$$\begin{aligned}A_{ii} &= (-w_i + u_{i-1})x, \\ B_{ii} &= (w_i + u_{i-1})x^2.\end{aligned} \quad (17)$$

Eq. (16) is the Fokker Planck equation associated with the MC method, and written in matrix form. Eq. (9) serves as the matrix solution of this Fokker Planck equation. From Eq. (17) and the Glauber transition rate [Eq. (7)], and after Taylor expansion of $p_{i+1}$ and $p_{i-1}$ about $x_i$ (such that the odd term vanishes), we obtain

$$B_{ii} = (w_i + u_{i-1})x^2 = x^2 + O(x^4) \quad (18)$$

Thus, we can write the FP diffusion coefficient for the MC method as

$$\begin{aligned}B_{MC} &= \lim_{\Delta t \to 0}\frac{1}{\Delta t}\langle \Delta x^2\rangle = \langle B_{ii}\rangle \\ &= x^2\end{aligned} \quad (19)$$

In the above, we have chosen $\Delta t$ to represent the MC time step, to differentiate from the real time step $\Delta t$ of the LD method. Thus, we have expressed both the LD equation and the master equation of the MC method in the Fokker-Planck form. We can now make a direct comparison of the diffusion coefficient $B$ of these two Fokker-Planck equations [i.e. Eqs. (13) and (19)], and derive the relationship between MCS and the real time unit, i.e.

$$\langle \Delta x^2\rangle = \frac{2k_B T a g_0}{(1+a^2)m_s}\Delta t = x^2 \Delta t = \left(\frac{2}{N}\right)^2 \Delta t \quad (20)$$

Eq. (20) is the main result of time quantification of the RWMC method. Eq. (20) enables us to model in real time units the thermally induced magnetization dynamics of an isolated single domain magnetic particle by using explicit matrix solutions [e.g. Eq. (9)].

## IV. RESULTS AND DISCUSSION

To test the validity of Eq. (20), we investigate the magnetization reversal of an isolated single domain particle by both LD and MC methods. In the LD case, this process is modeled by direct time-step integration of the LD equation [Eq. (2)], while for the MC method, the simulation is performed based on Eq. (9), and the application of the conversion factor of Eq. (20). The results are plotted in Fig. 2, and it is clear that the LLG results and the numerical RWMC results agree very well, after the application of the conversion factor. We also plot the analytical approximations of the switching time based on the second largest eigenvalue $l_1$ and the "effective" eigenvalue $l_{eff}$, as well as the asymptotic analytical result by Brown *et al*18,[25],[26]. At low temperatures these analytical results are in very good agreement with both numerical LD and RWMC methods. At high temperature ($s<2$) however, a small divergence of ~ 10% to 15% occurs between the LD results and the analytical RWMC results. In this high temperature region, $l_{eff}^{-1}$ yields a much better convergence compared to the simpler approximation $l_1^{-1}$. As a further test of the relative accuracy of $l_1^{-1}$ and $l_{eff}^{-1}$ approximations, we compare their predictions to the actual numerical LD result for the oblique easy axis case ($q = p/4$). In this case, it becomes even more apparent that the $l_{eff}^{-1}$ approximation is more robust and capable of good accuracy for a larger range of temperatures and easy-axis orientations.

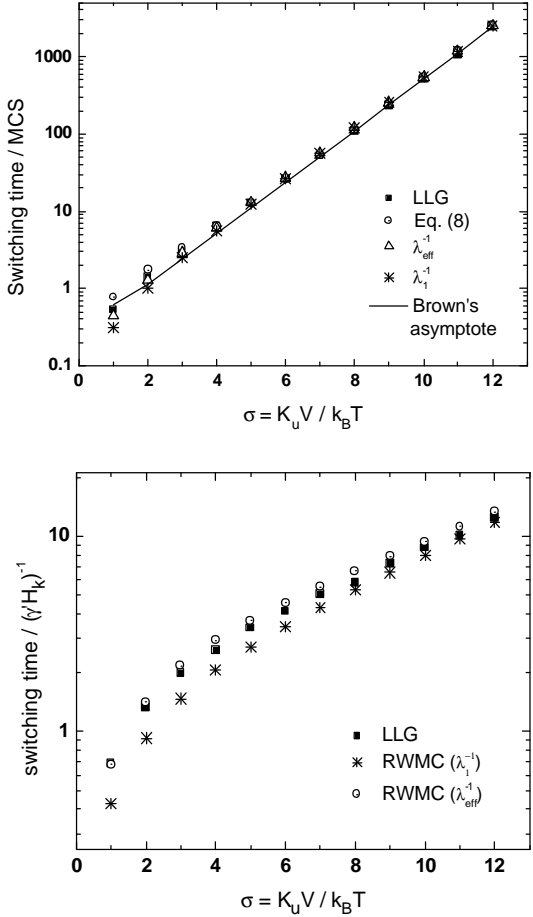

Fig. 2. Switching time vs. temperature, for a) $q = 0$, Applied field $h = -0.1$. Damping constant a = 4. b) $q = p/4$, Applied field $h = -0.22$. Damping constant a = 2.

As a statistical description of magnetization reversal, RWMC method is not able to describe the precessional dynamics during a reversal process. This is because precessional motion is an athermal process, which is essentially driven by the effective magnetic field and not by thermal fluctuations. Chubykalo *et al.* investigated the conditions under which the influence of precession becomes significant and leads to the breakdown in the MC approximation[9]. We confirm this finding in the time quantification RWMC method by investigating the dependence of the switching time on the damping parameter *a*. As seen in Fig. 3, in the symmetric case (uniaxial single domain particle), the precessional motion does not affect the accuracy of the MC method, while in the nonsymmetric oblique case, the RWMC method gives the accurate result only at high damping condition, where precessional motion is suppressed.

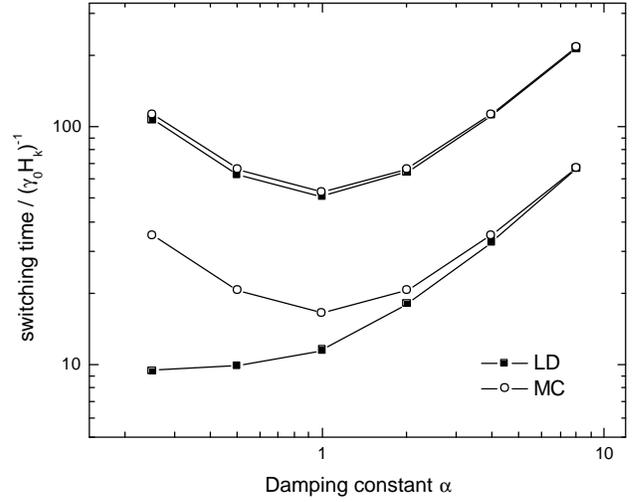

Fig. 3. Switching time vs. damping constant a for $K_uV/k_BT = 4.5$. A field $h = -0.15$ applied under an angle of 0 (top) and $p/4$ to the easy axis (bottom).

## V. CONCLUSION

We have applied the RWMC simulation method to model the thermally induced magnetization dynamics of a Brownian single domain particle model. By considering the alternative LD method, and directly comparing the corresponding Fokker-Planck equations of both methods, we derive an analytical conversion factor between MCS unit into real physical time. This time quantification is verified by the close agreement between the time-quantified MC results and LD numerical data. We also derive an analytical expression of the switching time (based on an "effective" eigenvalue), which goes beyond the usual approximation based on the second largest eigenvalue $l_1$. We compare the predictions of both analytical approximations to the numerical LD and RWMC results and found that the effective eigenvalue approximation shows more robustness especially in the high temperature and oblique easy axis cases, for which the simpler $l_1$ approximation breaks down. Finally, we provide a second validation of the time quantification by examining the influence of the damping constant parameter *a* on the switching time. We obtain a convergence of both time quantified RWMC with LD results at all *a* for the symmetric case. However, a divergence occurs at low *a* in the oblique case (due to the breakdown of the RWMC method in modeling precessional modes), which is in agreement with previous works.

## Appendix (I)

The transition matrix of a birth-death process has the form of

$$\mathbf{A}_{n \times n} = \begin{pmatrix} -w_1 & u_1 & & & & \\ w_1 & -w_2-u_1 & u_2 & & & \\ & w_2 & \cdots & & & \\ & & & \cdots & u_{n-1} \\ & & & w_{n-1} & -u_{n-1} \end{pmatrix}.$$

We would like to derive an analytical estimate of the second largest eigenvalue. (All eigenvalues are negative except for the largest $l_0 = 0$).

### A)  *Properties of determinant*

Let the polynomial $J_n(x) = |x\mathbf{I} - \mathbf{A}|$, namely

$$J_n(x) = \begin{vmatrix} x+w_1 & u_1 & & & & \\ w_1 & x+w_2+u_1 & u_2 & & & \\ & w_2 & \cdots & & & \\ & & & \cdots & u_{n-1} \\ & & & w_{n-1} & x+u_{n-1} \end{vmatrix}.$$

We define another polynomial $K_n(x)$:

$$K_n(x) = \begin{vmatrix} x+w_1 & u_1 & & & & \\ w_1 & x+w_2+u_1 & u_2 & & & \\ & w_2 & \cdots & & & \\ & & & \cdots & u_{n-1} \\ & & & w_{n-1} & x+u_{n-1}+w_n \end{vmatrix}.$$

The only difference between $J_n(x)$ and $K_n(x)$ is the extra $w_n$ term for the bottom right corner element of the determinant. Thus, $J_n(x)$ and $K_n(x)$ have the relationship as:

$$K_n = J_n + w_n K_{n-1} \qquad (a1)$$

On the other hand, the definition of determinant yields another relationship $J_n(x)$ and $K_n(x)$ as:

$$J_n = (x + u_{n-1})K_{n-1} - w_{n-1} u_{n-1} K_{n-2} \qquad (a2)$$

From Eqs. (a1) and (a2), we have

$$\begin{aligned} J_n &= (x + u_{n-1})K_{n-1} - w_{n-1} u_{n-1} K_{n-2} \\ &= (x + u_{n-1})K_{n-1} - u_{n-1}(K_{n-1} - J_{n-1}) \end{aligned}$$

$$J_n = x K_{n-1} + u_{n-1} J_{n-1} \qquad (a3)$$

Substituting Eq. (a1) into Eq. (a3), we obtain

$$\begin{aligned} J_n &= x(J_{n-1} + w_{n-1} K_{n-2}) + u_{n-1} J_{n-1} \\ &= (x + u_{n-1})J_{n-1} + w_{n-1}(x K_{n-2}) \\ &= (x + u_{n-1})J_{n-1} + w_{n-1}(J_{n-1} - u_{n-2} J_{n-2}), \end{aligned}$$

where we have made use of Eq. (a3) in the last step of the above derivation. So we obtain the following difference equation:

$$J_n = (x + u_{n-1} + w_{n-1})J_{n-1} - w_{n-1}u_{n-2}J_{n-2}, \qquad (a4)$$

with the initial conditions

$$J_0 = 0,$$
$$J_1 = x.$$

$J_n(x)$ is an $n$-th order polynomial and it is straightforward to see that $x_0 = 0$ is a root of $J_n(x)$ (n.b. $J_0$, $J_1$ being proportional to $x$). In particular, with the assumption of Glauber's rate, i.e. $w_i + u_i = 1$, $J_n(x)$ can be simplified to be:

$$J_n(x) = x\left[\sum_{i=1}^{n-1} A_i \cdot (1+x)^i\right]$$

***B) The first order approximation of second largest eigenvalue.***

Here we define another polynomial function $f_n(x)$:

$$f_n(x) = J_n(x)/x. \qquad (a5)$$

We would like to derive the approximation $x_{eff}$ of the largest nonvanishing root $x_1$ of $f_n(x)$ using first-order approximation which is given by:

$$x_{eff} = -f_n(0)/f_n'(0) \qquad (a6)$$

From Vieta's theorem we know that $x_{eff} = \left(x_1^{-1} + x_2^{-1} + \cdots + x_{n-1}^{-1}\right)^{-1}$ which includes the contribution from all roots of Eq. (a5). This first-order approximation will yield good accuracy under the condition of $|x_1| << |x_i|$ $(i > 1)$.

We will now calculate $f_n(0)$ and $f_n'(0)$ respectively. To calculate $f_n(0)$, we consider Eqs. (a4) and (a5), from which we have

$$f_n(x) = (x + u_{n-1} + w_{n-1}) f_{n-1}(x) - w_{n-1} u_{n-2} f_{n-2}(x) \qquad (a7)$$

and the initial conditions:

$$f_0(x) = 0,$$
$$f_1(x) = 1.$$

When $x = 0$, Eq. (a7) results in

$$f_n(0) = (u_{n-1} + w_{n-1}) f_{n-1}(0) - w_{n-1} u_{n-2} f_{n-2}(0),$$

so that

$$f_n(0) - u_{n-1}f_{n-1}(0) = w_{n-1}(f_{n-1}(0) - u_{n-2}f_{n-2}(0))$$
$$= \prod_{i=1}^{n-1} w_i. \quad (a8)$$

Let

$$g_n = f_n(0) / \prod_{i=1}^{n-1} u_i, \quad (a9)$$

and let us define a series $\{p_i\}$ such that:

$$p_1 = 1$$

$$p_i = \prod_{j=1}^{i-1} \frac{w_j}{u_j} \quad (1 < i \leq n). \quad (a10)$$

Eq. (a8) thus reduces to

$$g_n - g_{n-1} = p_n,$$

so that

$$g_n = \sum_{i=1}^{n} p_i.$$

We can thus calculate $f_n(0)$:

$$f_n(0) = \left(\prod_{i=1}^{n-1} u_i\right) g_n = \prod_{i=1}^{n-1} u_i \cdot \left(\sum_{j=1}^{n} p_j\right) \tag{a11}$$

We would like to mention that the series of {$p_i$} defined above also have a physical meaning, i.e. the *relative probability distribution at equilibrium.* [a1]

The calculation of $f_n'(0)$ is more involved and the full derivation is given in appendix (II). The result of $f_n'(0)$ is given as:

$$f_n'(0) = \left(\prod_{i=1}^{n-1} u_i\right) \cdot \sum_{i=1}^{n-1} \left(\frac{1}{p_{i+1} u_i} \left(\sum_{j=1}^{i} p_j\right)\left(\sum_{k=i+1}^{n} p_k\right)\right). \tag{a12}$$

So that from Eq. (a6), we obtain $x_{eff}$:

$$x_{eff} = -f_n(0)/f_n'(0) = -\frac{\sum_{i=1}^{n} p_i}{\sum_{i=1}^{n-1}\left(\frac{1}{p_{i+1} u_i}\left(\sum_{j=1}^{i} p_j\right)\left(\sum_{k=i+1}^{n} p_k\right)\right)} \tag{a13}$$

If Glauber rate is used for $u_i$, we will then have

$$x_{eff} = -\frac{\sum_{i=1}^{n} p_i}{\sum_{i=1}^{n-1}\left(\left(\frac{1}{p_i} + \frac{1}{p_{i+1}}\right)\left(\sum_{j=1}^{i} p_j\right)\left(\sum_{k=i+1}^{n} p_k\right)\right)} \tag{a14}$$

*C)   Accuracy of the approximation.*

To examine the accuracy of the approximation, we consider a model probability distribution of

$$p(x) = C \exp(s \cdot x^2).$$

The bin number is set to be $N = 64$.

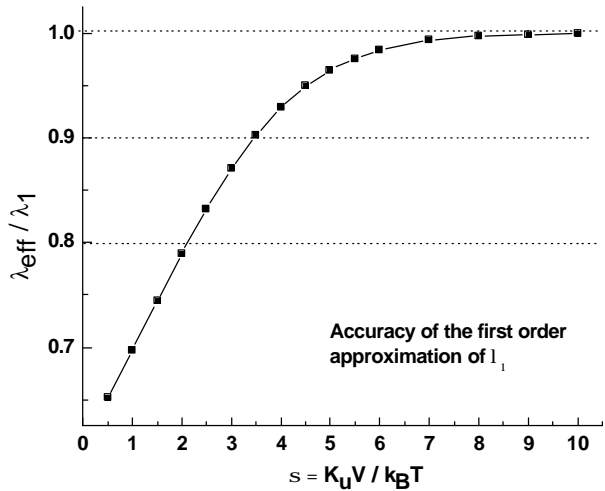

**Fig. A1. Plot of the accuracy as $l_{eff}/l_1$ versus $s$.**

[a1] Samuel Karlin and Howard M. Taylor, *A first course in stochastic processes*, **2nd edition**, Academic press, New York.

**Appendix (II):**

For a given function

$$f_n(x) = (x + u_{n-1} + w_{n-1})f_{n-1}(x) - w_{n-1}u_{n-2}f_{n-2}(x) \qquad (b1)$$

$$f_0(x) = 0,$$
$$f_1(x) = 1;$$

and

$$f_n(0) = \prod_{i=1}^{n-1} u_i \cdot \left( \sum_{j=1}^{n} p_j \right), \qquad (b2)$$

we want to obtain $f_n'(0)$.

Since

$$f_n'(0) = f_{n-1}(0) + (u_{n-1} + w_{n-1})f_{n-1}'(0) - w_{n-1}u_{n-2}f_{n-2}'(0), \qquad (b3)$$

and using the same definition as Eq. (a9),

$$g_n = f_n(0) / \prod_{i=1}^{n-1} u_i = \sum_{i=1}^{n} p_i, \qquad g_n' = f_n'(0) / \prod_{i=1}^{n-1} u_i$$

Eq. (b3) can be reduced to:

$$g_n' - g_{n-1}' = \frac{g_{n-1}}{u_{n-1}} + \frac{w_{n-1}}{u_{n-1}}(g_{n-1}' - g_{n-2}')$$

$$= \frac{g_{n-1}}{u_{n-1}} + \frac{p_n}{p_{n-1}}(g_{n-1}' - g_{n-2}')$$

$$= \frac{p_n}{p_n} \frac{g_{n-1}}{u_{n-1}} + \frac{p_n}{p_{n-1}} \frac{g_{n-2}}{u_{n-2}} + \cdots + \frac{p_n}{p_2} \frac{g_1}{u_1}$$

In above derivation, $\dfrac{p_n}{p_{n-1}} = \dfrac{w_{n-1}}{u_{n-1}}$ is used according to Eq. (a10). Let

$$k_n = g_n' - g_{n-1}',$$

so that:

$$g_n' = k_n + g_{n-1}' = \sum_{i=1}^{n} k_i.$$

For clarity, we expand the expression of $k_n$, $k_{n-1}$ and $k_{n-2}$ etc.

$$k_n = \frac{p_n}{p_n}\frac{g_{n-1}}{u_{n-1}} + \frac{p_n}{p_{n-1}}\frac{g_{n-2}}{u_{n-2}} + \frac{p_n}{p_{n-2}}\frac{g_{n-3}}{u_{n-3}} + \cdots + \frac{p_n}{p_2}\frac{g_1}{u_1},$$

$$k_{n-1} = \frac{p_{n-1}}{p_{n-1}}\frac{g_{n-2}}{u_{n-2}} + \frac{p_{n-1}}{p_{n-2}}\frac{g_{n-3}}{u_{n-3}} + \cdots + \frac{p_{n-1}}{p_2}\frac{g_1}{u_1},$$

$$k_{n-2} = \frac{p_{n-2}}{p_{n-2}}\frac{g_{n-3}}{u_{n-3}} + \cdots + \frac{p_{n-2}}{p_2}\frac{g_1}{u_1},$$

so that,

$$g_n' = \sum_{i=1}^{n} k_i = f_n'(0) = \sum_{i=1}^{n-1}\left(\frac{1}{p_{i+1}u_i} g_i \left(\sum_{k=i+1}^{n} p_k\right)\right).$$

Finally we obtain Eq. (a12), i.e.

$$f_n'(0) = \left(\prod_{i=1}^{n-1} u_i\right) \cdot g_n' = \left(\prod_{i=1}^{n-1} u_i\right) \cdot \sum_{i=1}^{n-1}\left(\frac{1}{p_{i+1}u_i}\left(\sum_{j=1}^{i} p_j\right)\left(\sum_{k=i+1}^{n} p_k\right)\right).$$